\documentclass[11pt]{article}
\pdfoutput=1
\usepackage[utf8]{inputenx}
\usepackage{jheppubmod}
\usepackage{pstool}
\PassOptionsToPackage{normalem}{ulem}
\usepackage{ulem}
\usepackage{braket}
\usepackage{bbm}
\usepackage{multirow}
\usepackage{enumitem}
\usepackage{simplewick}
\usepackage{physics}
\usepackage{graphicx}
\usepackage[font=small]{caption}
\usepackage{subcaption}

\usepackage{xcolor}

\newcommand{\normord}[1]{:\mathrel{#1}:}
\allowdisplaybreaks
\pdfoutput=1

\definecolor{DarkGreen}{RGB}{18,173,42}
\definecolor{DarkRed}{RGB}{202,0,42}
\definecolor{Gray}{gray}{0.5}

\DeclareMathAlphabet{\mathbbold}{U}{bbold}{m}{n}

\def\bfone{\relax{\rm 1\kern-.35em 1}}

\newcommand{\noprint}[1]{}

\newcommand{\be}{\begin{equation}}
\newcommand{\ee}{\end{equation}}
\newcommand{\ben}{\begin{displaymath}}
\newcommand{\een}{\end{displaymath}}
\newcommand{\bea}{\begin{eqnarray}}
\newcommand{\eea}{\end{eqnarray}}

\newcommand{\bean}{\begin{eqnarray*}}
\newcommand{\eean}{\end{eqnarray*}}

\DeclareMathAlphabet{\mathpzc}{OT1}{pzc}{m}{it}

\title{Quenched coupling, entangled equilibria, and correlated composite operators: a tale of two $O(N)$ models}

\author[a]{Souvik Banerjee,}
\emailAdd{souvik.banerjee@uu.se}
\affiliation[a]{Department of Physics and Astronomy, Uppsala University, \\
 SE-751 08 Uppsala, Sweden.\\ }

\author[b]{Julius Engelsöy,}
\emailAdd{julius.engelsoy@fysik.su.se}

\author[b]{Jorge Larana-Aragon,}
\emailAdd{jorge.laranaaragon@fysik.su.se}

\author[b]{Bo Sundborg,}
\emailAdd{bo@fysik.su.se}
\affiliation[b]{The Oskar Klein Centre for Cosmoparticle Physics \& Department of Physics, Stockholm University, \\
AlbaNova, 106 91 Stockholm, Sweden.\\}

\author[b,c]{\\ Larus Thorlacius,}
\emailAdd{larus.thorlacius@fysik.su.se}
\affiliation[c]{University of Iceland, Science Institute, \\ 
Dunhaga 3, 107 Reykjavik, Iceland\\}

\author[d]{Nico Wintergerst}
\emailAdd{nico.wintergerst@nbi.ku.dk}
\affiliation[d]{The Niels Bohr Institute, University of Copenhagen, \\
Blegdamsvej 17, 2100 Copenhagen Ø, Denmark\\}

\vspace{3mm}

\abstract{
A macroscopic version of Einstein--Podolsky--Rosen entanglement is obtained by quenching a quadratic coupling between two $O(N)$ vector models.  A quench of the mixed vacuum produces an excited entangled state, reminiscent of purified thermal equilibrium, whose properties can be studied analytically in the free limit of the individual field theories. The decoupling of different wavelength modes in free field theory prevents true thermalisation but a more subtle difference is that the density operator obtained by a partial trace does not commute with the post-quench Hamiltonian. Generalized thermal behaviour is obtained at late times, in the limit of weak initial mixing or a smooth but rapid quench. More surprisingly, late-time correlation functions of composite operators in the post-quench free field theory share interesting properties with correlators in strongly coupled systems. We propose a holographic interpretation of our result.}

\preprint{\parbox{3cm}{UUITP-10/19}}
\begin{document}
\maketitle

\section{Introduction}

Coupled fields and entanglement are ubiquitous in quantum field theory but it is challenging to get a handle on the entanglement structure of an interacting macroscopic system. 
In the thermofield double formulation, thermal field theories or black holes are represented by very special entangled states. In light of this, previous work on thermal free field theory and black holes \cite{AmadoSundborgThorlaciusWintergerst2017,AmadoSundborgThorlaciusWintergerst2018} suggests that specially entangled states of free field theory can mimic effects of strongly coupled field theory. Since these indications are somewhat surprising, it is important to check whether they are due to very fine-tuned entanglement. Requiring also that it is posssible to prepare the entangled state dynamicallly, we are led to consider entanglement generated by a quantum quench.
The construction involves a pair of quantum fields that are initially uniformly coupled at all points in space but otherwise free. The coupling between the fields is quenched and the subsequent evolution of the decoupled, but entangled, free field theories is then followed.

The physics is similar to that seen in homogenous quantum quenches in quantum field theories \cite{CalabreseCardy2006}, but the explicit entanglement between subsystems brings in new features. The late time behaviour of reduced density operators and correlation functions long after the quench is approximately thermal in certain limits, but deviations from thermal behaviour appear both in the decoupling of modes of a free field and from off-diagonal terms in the modular Hamiltonian obtained for each mode. A quench of mixing between two interacting subsystems would be expected to yield a purified double of a thermal state while in our simple system we have to settle for a state in a generalised Gibbs ensemble. 

The novelty of our approach is twofold. First, we probe the dynamic nature of the entanglement generation from a local relativistic Hamiltonian.
	 In this regard, our approach differs substantially from engineering a highly finetuned non-local Hamiltonian whose ground state is the thermofield double state \cite{CottrellFreivogelHofmanLokhande2018}. It can be viewed as a generalization to states with a less restricted entanglement structure. Second, we can probe consequences of entanglement that are independent of the coupling within the individual field theories.
Remarkably,
 some of our findings for free fields are similar to results that have been obtained at intermediate or strong coupling. Our quench protocol is analogous to the recently studied hard quench of interactions between two quantum-mechanical Sachdev-Ye-Kitaev (SYK) models \cite{MaldacenaQi2018}.\footnote{In fact, the SYK quench of \cite{MaldacenaQi2018} and the quench considered in this paper are both modelled on the original EPR quench \cite{EinsteinPodolskyRosen1935}.} The SYK model is of course quite different from a weakly interacting local quantum field theory, having more in common with the strong-coupling physics of holographic systems, 
yet
 we see some parallel behaviour.

We find, in particular, that a quench of coupled free fields yields qualitatively similar results to a quench at strong coupling, when we study \emph{composite operators}.  An instantaneous quench is imprinted on response functions of composite operators long after the quench.
The quench induces oscillations, but even after these have subsided, composite operators respond to disturbances at spacelike relative momenta. Such correlations are forbidden in the vacuum of a relativistic field theory, but here they remain imprinted in the medium arbitrarily late after the quench.

For a strongly coupled field theory, this effect has a curious holographic interpretation in terms of so called evanescent modes, which are supported in the vicinity of black hole horizons and decay exponentially radially outwards from the horizon \cite{ReyRosenhaus2014}. Comparison with this holographic result makes it clear that our example of entangled free subsystems actually displays features which are commonly attributed to strongly interacting systems.
	 This opens up the possibility to model strongly interacting systems by
simple entanglement of very simple building blocks.

Our observation of evanescent modes after a quench even at weak coupling may be interpreted in the form of a general ER = EPR relation \cite{MaldacenaSusskind2013} without requiring an explicit semiclassical bulk description. 
	The entanglement which is a dynamical consequence of the setup leads to bulk evanescent modes detectable in each decoupled field theory. Entanglement is also expected to be a minimal condition for the emergence of a smooth bulk geometry \cite{Van-Raamsdonk2010}. Since evanescent modes indicate bulk horizons, a simple bulk interpretation presents itself. An Einstein-Rosen bridge connecting two asymptotically AdS regions, one for each field theory. Except for this brief sketch, we defer a holographic analysis to Section \ref{sect:interpretation}.

The presence of spacelike correlations has been seen previously in certain large~$N$ thermal field theories, even in the limit of vanishing coupling \cite{JevickiYoon2016,JevickiSuzuki2016,AmadoSundborgThorlaciusWintergerst2018}. 
That different momentum modes thermalise independently after the quench however means that recent bounds \cite{BanerjeePapadodimasRajuSamantrayShrivastava2019} on correlations at spacelike momenta, obtained from thermal field theory, 
are not applicable. In fact, in contrast to the thermal case, correlators after the instantaneous quench are not exponentially damped at large momenta.
However, softening the quench from instantaneous to smooth improves the approximate fit to a thermal state and introduces a temperature scale inversely related to the time scale of the quench. 

We consider a system of two free $O(N)$ models, each containing $N$ free scalars related by a global $O(N)$ symmetry. This is the simplest large $N$ system which can accommodate the quench physics we are interested in. As further motivation, we note that a closely related theory, the free massless $O(N)$ model, is conjectured to have a higher spin gravity dual \cite{KlebanovPolyakov2002}. The observed connections between doubled $O(N)$ models and evanescent modes \cite{JevickiYoon2016,JevickiSuzuki2016} are intriguing to explore in a standard field theory setting. The instantaneous quench amounts to suddenly removing an initially present local coupling between the two $O(N)$ models, at a time chosen to be $t=0$ without loss of generality. In addition to its intrinsic interest, the quench permits us to investigate if a state supporting interesting modes with spacelike momenta can arise dynamically. 

The paper is organised as follows. In Section \ref{sect:mixing} we set up the mixed system, the quench and much of our notation. Two technical sections follow. Section \ref{sect:density_operators} describes the quantum state of the system in a language appropriate after the quench, and especially how it appears if reduced to one of the decoupled fields. Similarities and differences to a thermal description are probed and we observe that a smooth but fast quench provides a better approximation to a thermal state than an instantaneous quench. In Section \ref{sect:dissipation} we calculate two-point functions of composite operators, in particular dissipative contributions to retarded correlators. This is where results reminiscent of strong coupling physics are obtained, as discussed in Section \ref{sect:interpretation}. 
Finally, our main conclusions are summarised in Section~\ref{sect:conclusions}. Some technical details are referred from the main text to the Appendices. 

\section{Mixing and quench of mixing}\label{sect:mixing}

We wish to study the effect of suddenly decoupling fields which are initially mixed. We thus consider a system of two $O(N)$ vector models, described by fields $\varphi_L$ and $\varphi_R$, with a mass mixing that is turned off at time $t = 0$. 
The action is
 \begin{align}\label{eq:action}
 S &= \frac{1}{2}\int \mathrm{d}^{d}x \bigg[\partial_{\mu}\varphi^{(a)}_L\partial^{\mu}\varphi_L^{(a)} + \partial_{\mu}\varphi_R^{(a)}\partial^{\mu}\varphi_R^{(a)} - m_L^{2}\varphi_L^{(a)}\varphi_L^{(a)}- m_R^{2}\varphi_R^{(a)}\varphi_R^{(a)} - 2h^{2}(t)\varphi_L^{(a)}\varphi_R^{(a)}\bigg]\,.
 \end{align}
``$(a)$'' is the $O(N)$ index which from what follows we will drop to simplify our notation.
For an instantaneous quench, focus of our interest in most parts of this work, one has $h^{2}(t) = h^2\, \theta (-t)$. 
At $t<0$, because of the non-zero coupling, only the symmetry corresponding to the diagonal part of the full $O(N)_L\times O(N)_R$ symmetry of two decoupled theories is manifest in this action. In fact, the full symmetry breaking pattern can be understood as follows. For equal masses, $m_L = m_R$, and no mixing, $h = 0$, the action \eqref{eq:action} is invariant under $O(2N)$. A mass difference breaks this symmetry down to $O(N)_L \times O(N)_R$, while a non-zero $h$ breaks it down to a \emph{different} $O(N)_+ \times O(N)_-$. Only the diagonal parts of these subgroups agree and are unbroken in either case. In this sense the theory before the quench breaks the post-quench $O(N)_L \times O(N)_R$ down to $O(N)_\text{diag}$.

The action in Eq.\ \eqref{eq:action} can be diagonalized by defining rotated fields $\chi^\pm$
\begin{equation}
\label{eq:o2rotation}
\begin{pmatrix}
\varphi_L \\ \varphi_R
\end{pmatrix} =  \begin{pmatrix}
\cos\alpha & \quad
-\sin\alpha \\
\sin\alpha & \quad
\cos\alpha
\end{pmatrix} \begin{pmatrix}
\chi^+\\ \chi^-
\end{pmatrix}\,,
\end{equation}
with
$\tan2\alpha \equiv \frac{2h^2}{\Delta m^2}$
and $\Delta m^2 \equiv m_L^{2}-m_R^{2}$.

The action for the rotated fields takes the form
\begin{equation}\label{eq:new_action}
S = \frac{1}{2}\int \mathrm{d}^{d}x \bigg[\left(\partial_{\mu}\chi^-\right)^2 - m_-^{2}(\chi^-)^2\bigg] + \frac{1}{2}\int \mathrm{d}^{d}x \bigg[\left(\partial_{\mu}\chi^+\right)^2 - m_+^{2}(\chi^+)^2\bigg]\,.
\end{equation}
with
\begin{equation}
m_\pm^{2} = \frac{1}{2}\left(m_L^{2}+m_R^{2}\pm\sqrt{4h^{4}+(\Delta m^2)^{2}}\right)\,.
\end{equation}
This is the action that describes the complete pre-quench dynamics.

In the remainder of the paper, we will consider the special case $m_R = m_L = m$. Correspondingly, $\Delta m^2 = 0$ and we obtain for the masses and the field relations
\begin{align}
m_\pm^{2} &= m^2 \pm h^2 \,\theta(-t)\,,\\
\chi^\pm &= \frac{1}{\sqrt{2}}\left(\varphi_R \pm \varphi_L\right)\,.
\label{eq:equalmass}
\end{align}
This considerably simplifies many calculations. Nonetheless, our conclusions also apply to the case of $m_L \neq m_R$.

Let us note here that in the case of equal masses, the action \eqref{eq:action} for $h = 0$ has the full $O(2N)$ symmetry that would in principle allow us to choose $\chi^\pm$ as a basis even after the quench. In this case, the physical picture is that of two simultaneous mass quenches, from $m^2 \pm h^2$ to $m^2$. However, we implicitly assume that there are observers associated with the individual fields $\varphi_L$ and $\varphi_R$, which can for example be made manifest by adding explicit sources to the action \eqref{eq:action}. Correspondingly, we also choose to derive all results explicitly in the basis of $\varphi_L$ and $\varphi_R$. The only exception is Section \ref{subsect:smooth}, where we use the above $O(2N)$ symmetry to make contact with existing results in the literature.

\subsection{Relation between quantum fields before and after the quench}

Before the quench, i.e., for $t<0$, the relevant mode expansion is in terms of the $\chi$-fields which we may expand in terms of momentum modes, which for $t\to0^{-}$ take the form:
\begin{align}
\label{eq:chi_decomp}
\chi^\pm &= \int\frac{\dd {\bf k}}{(2\pi)^{d-1}}\frac{1}{\sqrt{2\omega^\pm_{\bf k}}}e^{i{\bf k}\cdot {\bf x}}\left(a_{{\bf k}}^{\pm} + (a_{-{\bf k}}^{\pm})^{\dagger}\right)\,,\\
\pi_\chi^{\pm} &= \partial_{t}\chi^\pm = -i\int\frac{\dd {\bf k}}{(2\pi)^{d-1}}\sqrt{\frac{\omega^\pm_{\bf k}}{2}}e^{i{\bf k}\cdot {\bf x}}\left(a_{{\bf k}}^{\pm} - (a_{-{\bf k}}^{\pm})^{\dagger}\right)\,,
\end{align}
where $\omega^\pm_{\bf k} = \sqrt{|{\bf k}|^2+m_\pm^{2}}$.
On the other hand, just after the quench, or equivalently for $t\to 0^{+}$, the relevant mode expansion is in terms of the $\varphi$-fields,
\begin{align}
\label{eq:phi_decomp}
&\varphi_L = \int\frac{\dd {\bf k}}{(2\pi)^{d-1}}\frac{1}{\sqrt{2\omega_{\bf k}}}e^{i{\bf k}\cdot {\bf x}}\left(L_{{\bf k}} + L_{-{\bf k}}^{\dagger}\right)\,,&\varphi_R = \int\frac{\dd {\bf k}}{(2\pi)^{d-1}}\frac{1}{\sqrt{2\omega_{\bf k}}}e^{i{\bf k}\cdot {\bf x}}\left(R_{{\bf k}} + R_{-{\bf k}}^{\dagger}\right)\,,\\
&\pi_L = -i\int\frac{\dd {\bf k}}{(2\pi)^{d-1}}\sqrt{\frac{\omega_{\bf k}}{2}}e^{i{\bf k}\cdot {\bf x}}\left(L_{{\bf k}} - L_{-{\bf k}}^{\dagger}\right)\,, &\pi_R = -i\int\frac{\dd {\bf k}}{(2\pi)^{d-1}}\sqrt{\frac{\omega_{\bf k}}{2}}e^{i{\bf k}\cdot {\bf x}}\left(R_{{\bf k}} - R_{-{\bf k}}^{\dagger}\right)\,,
\end{align}
where now $\omega_{\bf k} = \sqrt{|{\bf k}|^2+m^{2}}$. 

We impose the matching conditions
\begin{align}
\varphi_L &= \frac{1}{\sqrt{2}}\left(\chi^+ - \chi^-\right)\,,\quad
\varphi_R = \frac{1}{\sqrt{2}}\left(\chi^+ + \chi^-\right)\,,\\
\pi_L &= \frac{1}{\sqrt{2}}\left(\pi_\chi^{+} - \pi_\chi^{-}\right)\,,\quad \pi_R = \frac{1}{\sqrt{2}}\left(\pi_\chi^{+} + \pi_\chi^{-}\right)\,.
\end{align}
These can be solved to yield the Bogolyubov transformations
\begin{align}
\label{eq:bog}
L_{{\bf k}} 
&= \frac{1}{\sqrt{2}}\sum_{\sigma = \pm} u^{L\sigma}_{\bf k} a^{\sigma}_{\bf k} + \left(v^{L\sigma}_{\bf k}\right)^*(a^{\sigma}_{-{\bf k}})^{\dagger}\,,\\
R_{{\bf k}} 
&= \frac{1}{\sqrt{2}}\sum_{\sigma = \pm}u^{R\sigma}_{\bf k} a^{\sigma}_{\bf k} + \left(v^{R\sigma}_{\bf k}\right)^*(a^{\sigma}_{-{\bf k}})^{\dagger}\,,
\end{align}
The Bogolyubov coefficients are given by
\begin{equation}
\label{eq:bogcou}
u_{\bf k} = \frac{1}{2}
\begin{pmatrix}
\sqrt{\rule{0pt}{3.4ex}\frac{\omega^{}_{\bf k}}{\omega^{+}_{\bf k}}} + \sqrt{\frac{\omega^{+}_{\bf k}}{\omega^{}_{\bf k}}} &\quad - \left(\sqrt{\rule{0pt}{3.4ex}\frac{\omega^{}_{\bf k}}{\omega^{-}_{\bf k}}} + \sqrt{\frac{\omega^{-}_{\bf k}}{\omega^{}_{\bf k}}}\right) \\
\sqrt{\rule{0pt}{3.4ex}\frac{\omega^{}_{\bf k}}{\omega^{+}_{\bf k}}} + \sqrt{\frac{\omega^{+}_{\bf k}}{\omega^{}_{\bf k}}} &\quad \sqrt{\rule{0pt}{3.4ex}\frac{\omega^{}_{\bf k}}{\omega^{-}_{\bf k}}} + \sqrt{\frac{\omega^{-}_{\bf k}}{\omega^{}_{\bf k}}}
\end{pmatrix}\,,
\end{equation}
\begin{equation}
\label{eq:bogcov}
{v}_{\bf k} = \frac{1}{2}
\begin{pmatrix}
\sqrt{\rule{0pt}{3.4ex}\frac{\omega^{}_{\bf k}}{\omega^{+}_{\bf k}}} - \sqrt{\frac{\omega^{+}_{\bf k}}{\omega^{}_{\bf k}}} &\quad - \left(\sqrt{\rule{0pt}{3.4ex}\frac{\omega^{}_{\bf k}}{\omega^{-}_{\bf k}}} - \sqrt{\frac{\omega^{-}_{\bf k}}{\omega^{}_{\bf k}}}\right) \\
\sqrt{\rule{0pt}{3.4ex}\frac{\omega^{}_{\bf k}}{\omega^{+}_{\bf k}}} - \sqrt{\frac{\omega^{+}_{\bf k}}{\omega^{}_{\bf k}}} &\quad \sqrt{\rule{0pt}{3.4ex}\frac{\omega^{}_{\bf k}}{\omega^{-}_{\bf k}}} - \sqrt{\frac{\omega^{-}_{\bf k}}{\omega^{}_{\bf k}}}
\end{pmatrix}\,,
\end{equation}
where rows of the matrices are labelled by $L, R$ and columns by $+,-$.
As one can easily check, the respective matrix elements obey 
\begin{equation}
|u|^2 - |v|^2 = 1\,,
\end{equation}
thereby guaranteeing  canonical commutation relations for all creation and annihilation operators. Moreover, as evident from the above equations, the Bogolyubov coefficients for the instantaneous quench turn out to be real functions of the modulus of ${\bf k}$.

\section{Signs of thermality and non-thermality}\label{sect:density_operators}

Just after the quench, at $t = 0$, the system is still found in the vacuum of the coupled theories. However, in terms of the new Hamiltonian, this state corresponds to some highly excited state with nontrivial time dependence. It can be described either as a pure \emph{squeezed} state of the fields $\varphi_{L,R}$, or, by integrating out one of the latter, as a density matrix of $\varphi_L$ or $\varphi_R$ modes alone.

The nature of the density matrix depends on details of the quench. For example, quenches in simple two-mode systems can give rise to so-called two-mode squeezed states of the form (e.g. \cite{SchumakerCaves1985})
\begin{equation}
|\psi\rangle = e^{\zeta^* a b - \zeta a^\dagger b^\dagger} |0\rangle = \frac{1}{\cosh{|\zeta|}}\sum_{n=0}^\infty\left(-\frac{\zeta}{|\zeta|} \tanh{|\zeta|}\right)^n|n,n\rangle\,,
\end{equation}
where the parameter $\zeta$ is related to the strength of the quench and the state obeys a level matching condition $n_a = n_b$. 
The latter implies that $|\psi\rangle$ is a highly entangled state with a very particular entanglement structure. Tracing over one of the modes leads to an exactly thermal density matrix,
\begin{equation}
\rho_a = \frac{1}{\cosh^2{|\zeta|}}\sum_{n=0} \tanh^{2n}{|\zeta|} |n\rangle\langle n|\,,
\end{equation}
with an effective inverse temperature $\beta \sim -\log{\tanh^2{|\zeta|}}$.

As we will see, in our case the entanglement structure is more complicated. As a direct consequence of the relativistic invariance of the theory, the Hamiltonian at $t < 0$ mixes not only $L$ and $R$ modes, but also modes of wave vectors ${\bf k}$ and $-{\bf k}$. Upon tracing out e.g. the $R$-modes, this will lead to a density matrix for $L$ that is off-diagonal in the energy eigenbasis, with off-diagonal pieces parametrizing deviations from exact thermality (subsection \ref{subsect:Ensemble}).
Moreover, since we are dealing with a Gaussian field theory, with an infinite number of conserved charges, true thermalisation after the quench will not occur. At best, we can resort to the spirit of the generalized Gibbs ensemble \cite{RigolDunjkoYurovskyOlshanii2007}. There, one introduces a chemical potential for the conserved charges, of which in the case of free post-quench theories not all can be integrals over local functions of the fields \cite{EsslerMussardoPanfil2015,Sotiriadis2016}. 
 In accordance, we observe a momentum-dependent temperature (subsection \ref{subsect:CorrelatorTemperature}), similar to the case of mass quenched free fields \cite{CalabreseCardy2007}. The deviations from thermality vanish in the limit of weak coupling $h$ and small momentum $|{\bf k}|$.

The momentum dependence of the temperature is such that contributions to observables from high energy states are generally suppressed less than in ordinary thermal ensembles. In the case of an instantaneous quench, this holds up to arbitrarily high energies. In subsection \ref{subsect:smooth} we demonstrate that this can be cured by smoothing out the quench over a time scale $\Delta t$, in which case contributions from states with energy much greater than $1/\Delta t$ are in fact exponentially suppressed.

\subsection{The ensemble of modes decoupled by the quench}\label{subsect:Ensemble}

Directly after the quench, the system will find itself in a squeezed state of the modes $L$ and $R$. Correspondingly, the reduced density matrix of e.g. the $L$ modes will describe a mixed state whose phase factors will depend on the Bogolyubov coefficients. If the quenched state were an exact thermofield double state, the reduced density matrix would be diagonal. Generically, however, it will be described for a given wave vector ${\bf k}$ by 
\begin{equation}
\label{eq:densmat}
\rho_L^{\bf k} = {\cal N}_{\bf k} e^{-K_{\bf k}}\,,
\end{equation}
where the coefficient ${\cal N}_{\bf k}$ ensures unit normalization of the density matrix.
To construct $K$, we can make use of the fact that the pre-quench vacuum is a Gaussian state. Correspondingly, the density matrix of a single mode will also be Gaussian, with a modular Hamiltonian of the form \cite{FlassigPritzelWintergerst2013}
\begin{equation}
K_{\bf k} = A_{\bf k} \left(L_{\bf k}^\dagger L_{\bf k} + L_{\bf -k}^\dagger L_{\bf -k}\right) + B_{\bf k} L_{\bf k}^\dagger L_{-{\bf k}}^\dagger + B_{\bf k}^* L_{\bf k} L_{-{\bf k}}\,,
\end{equation}
where $O(N)$ indices are once again suppressed. The coefficient $A_{\bf k}$ is real. $B_{\bf k}$, generally complex, parametrizes the non-diagonal contributions to the density matrix and therefore deviations from thermality, $[H,K_{\bf k}] \sim B_{\bf k} L_{\bf k}^\dagger L_{-{\bf k}}^\dagger - B_{\bf k}^* L_{\bf k} L_{-{\bf k}}$. The coefficients can be fixed from low order correlation functions at $t = 0$, such as
$\langle L_{\bf k}^\dagger L_{\bf k} \rangle$ and $ \langle L_{\bf -k} L_{\bf k} \rangle$.
These are now calculated as usual via $\langle {\cal O}_{\bf k} \rangle = \tr ({\cal O}_{\bf k} \rho_L^{\bf k})$. Evaluation is simplest by performing a Bogolyubov rotation on the creation and annihilation operators $L$ in order to diagonalize the modular Hamiltonian. We define 
\begin{equation}
L_{\bf k} = r_{\bf k} c_{\bf k} + s_{\bf k}^* c_{-{\bf k}}^\dagger\,.
\label{eq:modbog}
\end{equation}
Making the ansatz $\arg(s_{\bf k}) = -\arg(r_{\bf k})$, inserting into the modular Hamiltonian for a given ${\bf k}$-mode and demanding diagonalization of $K_{\bf k}$ as well as the proper commutation relations, $[c_{\bf k},c_{\bf p}^\dagger] = (2\pi)^{d-1}\delta^{(d-1)}({\bf k - p})$, yields
\begin{align}
|r_{\bf k}|^2 &= \frac{1}{2}\left(1+\frac{A_{\bf k}}{\epsilon_{\bf k}}\right) \,,\\
|s_{\bf k}|^2 &= \frac{1}{2}\left(-1+\frac{A_{\bf k}}{\epsilon_{\bf k}}\right) \,,\\
\arg(r_{\bf k}) &= \frac{1}{2}\arg(B_{\bf k})\,,
\end{align}
with the one-particle energy $\epsilon_{\bf k}$ given by
\begin{equation}
\epsilon_{\bf k} \equiv \sqrt{A_{\bf k}^2 - |B_{\bf k}|^2}\,.
\end{equation}
Consequently, in the basis of $c_{\bf k}$, the modular Hamiltonian becomes
\begin{equation}
K_{\bf k} = \epsilon_{\bf k} \left(c_{\bf k}^\dagger c_{\bf k} + c_{\bf -k}^\dagger c_{\bf -k}\right)\,.
\end{equation}
We can now fix the normalization constant ${\cal N}_{\bf k}$, obtaining
\begin{equation}
1 = \tr (\rho_L^{\bf k}) = {\cal N}_{\bf k} \left(\sum_{m_+,m_-} e^{-\epsilon_{\bf k} (m_+ + m_-)}\right)^N = {\cal N}_{\bf k} \frac{1}{(1-e^{-\epsilon_{\bf k}})^{2N}}\,,
\end{equation}
where we take the trace in the number basis of $c_{\pm {\bf k}}$, with $m_\pm$ the eigenvalues of the respective number operator.

Inserting now the decomposition \eqref{eq:modbog} and observing that only number conserving operators can contribute, we obtain for the expectation values
\begin{align}
\frac{1}{N}\langle L_{\bf k}^\dagger L_{\bf k} \rangle &= {\cal N}_{\bf k} \sum_{m_+,m_-} e^{-\epsilon_{\bf k} (m_+ + m_-)}\left(|r_{\bf k}|^2 m_+ + |s_{\bf k}|^2(m_-+1)\right) = \frac{A_{\bf k}}{2\epsilon_{\bf k}}\coth(\frac{\epsilon_{\bf k}}{2})-\frac{1}{2}\,,\notag\\
\frac{1}{N}\langle L_{-{\bf k}} L_{\bf k} \rangle &= {\cal N}_{\bf k} r_{\bf k} s_{\bf k}^* \sum_{m_+,m_-} e^{-\epsilon_{\bf k} (m_+ + m_-)}\left(m_+ + m_-+1\right) = \frac{B_{\bf k}}{2\epsilon_{\bf k}}\coth(\frac{\epsilon_{\bf k}}{2})\,.
\label{eq:expvalL01}
\end{align}
As expected, $\langle L_{-{\bf k}} L_{\bf k} \rangle$, which is trivial in the $L$-vacuum, is directly proportional to $B_{\bf k}$.

All expectation values can now equivalently be calculated directly in the pre-quench vacuum state $|0\rangle_\chi$, making use of the Bogolyubov decomposition \eqref{eq:bog}. We obtain
\begin{align}
\frac{1}{N}\langle L_{\bf k}^\dagger L_{\bf k} \rangle &= \frac{1}{2}\sum_\sigma |v^{L\sigma}_{\bf k}|^2 \,,\notag\\
\frac{1}{N}\langle L_{-{\bf k}} L_{\bf k} \rangle &= \frac{1}{2}\sum_\sigma u^{L\sigma}_{\bf k}\left(v^{L\sigma}_{\bf k}\right)^*\,.
\label{eq:expvalL02}
\end{align}
Equating the corresponding expressions yields for the coefficients
\begin{align}
A_{\bf k} &= \frac{\sum_\sigma(u^{L\sigma}_{\bf k})^2 + (v^{L\sigma}_{\bf k})^2}{2{\cal F}_{\bf k}} \epsilon_{\bf k} = \frac{\epsilon_{\bf k}}{4{\cal F}_{\bf k}}\sum_\sigma\left(\frac{\omega_{\bf k}}{\omega_{\bf k}^\sigma}+\frac{\omega_{\bf k}^\sigma}{\omega_{\bf k}}\right)\,\notag
\,,\\
\label{eq:codm}
B_{\bf k} &= \frac{\sum_\sigma u^{L\sigma}_{\bf k}v^{L\sigma}_{\bf k}}{{\cal F}_{\bf k}} \epsilon_{\bf k} = \frac{\epsilon_{\bf k}}{4{\cal F}_{\bf k}}\sum_\sigma\left(\frac{\omega_{\bf k}}{\omega_{\bf k}^\sigma}-\frac{\omega_{\bf k}^\sigma}{\omega_{\bf k}}\right)\,,\\
\epsilon_{\bf k} &= 2\,\text{arccoth}({\cal F}_{\bf k})\notag\,,
\end{align}
where we have defined 
\begin{equation}
{\cal F}_{\bf k} \equiv \frac{1}{2}\sqrt{\sum_{\sigma,\sigma'}(u^{L\sigma}_{\bf k} + v^{L\sigma}_{\bf k})^2(u^{L\sigma'}_{\bf k} - v^{L\sigma'}_{\bf k})^2} = \frac{1}{2}\frac{\omega_{\bf k}^++\omega_{\bf k}^-}{\sqrt{\omega_{\bf k}^+ \omega_{\bf k}^-}}\,.
\end{equation}
We thus have derived an explicit expression for the modular Hamiltonian for the density matrix of $L$ modes in the pre-quench vacuum. All deviations from thermality of correlation functions of $\varphi_L$ are due to the off-diagonal terms in this modular Hamiltonian, quantified by the coefficient $B_{\bf k}$. 

We note that this expression allows us to directly compute the entanglement entropy between the $L$ and $R$-modes, defined as the von Neumann entropy of the density matrix \eqref{eq:densmat},
\begin{equation}
\label{eq:vNent}
S_L \equiv \frac{1}{2}\int \dd{\bf k}\,S_L^{\bf k} \equiv -\frac{1}{2}\int \dd{\bf k} \tr \left(\rho_L^{\bf k} \log \rho_L^{\bf k}\right)\,,
\end{equation}
where the factor of $\frac{1}{2}$ accounts for the fact that $\rho_L^{\bf k}$ involves both signs $\pm{\bf k}$.
Evaluated in the number basis of $c_{\pm\bf k}$, \eqref{eq:vNent} yields
\begin{equation}
S_L = N \int \dd{\bf k}\,\left[\frac{\epsilon_{\bf k}}{e^{\epsilon_{\bf k}}-1} - \log(1 - e^{-\epsilon_{\bf k}})\right] \,,
\end{equation}
which can be shown to agree perfectly with previous results, for example found in \cite{MollabashiShibaTakayanagi2014} for $N = 1$. We will not pursue further studies of entanglement entropy in this work.

As we will discuss in more detail in the ensuing sections, correlation functions of local operators become approximately time translationally invariant long after the quench. In that case, such late time observables can be obtained from an \emph{effective} ${\bf k}$-dependent thermal density matrix,
\begin{equation}
\label{eq:densdiag}
\rho^{\text{eff}}_{\bf k} = {\cal N}_{\bf k} e^{-\beta_{\bf k} \omega_{\bf k} L_{\bf k}^\dagger L_{\bf k}}\,,
\end{equation}
with a temperature that is fixed by relations \eqref{eq:expvalL01} and \eqref{eq:expvalL02}. In other words, one expects approximately thermal occupation
\begin{equation}
\langle L_{\bf k}^\dagger L_{\bf k}\rangle = \tr \left( L_{\bf k}^\dagger L_{\bf k}\,\rho^{\text{eff}}_{\bf k} \right) = \frac{1}{e^{\beta_{\bf k}\omega_{\bf k}} - 1}\,,
\end{equation}
which can be solved to give
\begin{equation}
\label{eq:temp_pred}
\beta_{\bf k} = \frac{1}{\omega_{\bf k}}\log{\left(\frac{\sum_\sigma (u_{\bf k}^{L\sigma})^2}{\sum_\sigma (v_{\bf k}^{L\sigma})^2}\right)}\,.
\end{equation}
We will confirm this expectation explicitly in section \ref{subsect:CorrelatorTemperature} below. 
The  ${\bf k}$-dependence of the temperature is natural, as discussed in the beginning of the section. The density matrix \eqref{eq:densdiag} can be reproduced from a generalized Gibbs ensemble,  constructed either directly from the momentum space charges $L_{\bf k}^\dagger L_{\bf k}$ or from a set of charges that is only mildly nonlocal \cite{Sotiriadis2016}. 

Let us note that in the limit of small $h$, we have up to order $h^4$ the following scaling of the parameters:
\begin{align}\label{eq:shbog}
\omega^\pm_{\bf k} &\approx \omega_{\bf k}\left(1 \pm \frac{h^2}{2\omega_{\bf k}^2} - \frac{h^4}{8\omega_{\bf k}^4}\right)\,,\notag\\
u^{L\sigma}_{\bf k} &= \sigma \left(1+\frac{h^4}{32 \omega_{\bf k}^4}\right)\,,\quad v^{L\sigma}_{\bf k} = \frac{1}{4}\frac{h^2}{\omega_{\bf k}^2}\left(-1+\sigma\frac{h^2}{2\omega_{\bf k}^2}\right)\,,
\end{align}
so that \eqref{eq:codm} becomes
\begin{equation}
\label{eq:smallhtemp}
A_{\bf k} = \epsilon_{\bf k}\,,\quad B_{\bf k} = \frac{h^4}{8 \omega_{\bf k}^4} \epsilon_{\bf k}\,,\quad\epsilon_{\bf k} = \log\frac{16\omega_{\bf k}^4}{h^4} - \frac{h^4}{16\omega_{\bf k}^4}\,.
\end{equation}
To leading order in $h$, the density matrix can be written as
\begin{equation}
\rho^{\bf k}_L \approx {\cal N}_{\bf k} e^{-\beta_{\bf k} H_{\bf k}}\,,\text{ with }H_{\bf k} = \omega_{\bf k} L_{\bf k}^\dagger L_{\bf k}\,,
\end{equation}
and
\begin{equation}
\label{eq:temperature}
\beta_{\bf k} \approx \omega_{\bf k}^{-1}\log\frac{16\omega_{\bf k}^4}{h^4}\,,
\end{equation}
which can be easily checked to agree with the leading order contribution to \eqref{eq:temp_pred}.
In passing, we note that the temperature $\beta_{\bf k}$ approaches a constant for long wavelengths. For these, we obtain an effectively thermal ensemble with
\begin{equation}
T \approx \frac{m}{\log{\frac{16m^4}{h^4}}}\,.
\end{equation}
As we can also see, small mixing leads to low temperature, consistent with the expectation that the effect of the quench decreases with decreasing coupling. Moreover, shorter wavelengths, with larger $|{\bf k}|$, will see a higher effective temperature. Correspondingly, while large $|{\bf k}|$ contributions to observables still decay, they do so slower than in an ordinary thermal ensemble, as a power law instead of exponentially. This, however, is caused by the instantaneous nature of the quench, as we demonstrate explicitly in subsection \ref{subsect:smooth}.

\subsection{The effective temperature from correlators}\label{subsect:CorrelatorTemperature}

As already mentioned above, one can define a generalized effective temperature by considering the large time behaviour of low order correlation functions of local operators \cite{CalabreseCardy2006,CalabreseCardy2007,SotiriadisCardy2008, SotiriadisCalabreseCardy2009}.  
We illustrate this using a simple correlator, namely the expectation value of a bilocal singlet operator in the pre-quench vacuum,
\begin{multline}
\langle\varphi_L({\bf x},t)\varphi_L({\bf x}',t')\rangle_{\chi} = \int\frac{\dd{\bf k}\,\dd{\bf k}'}{(2\pi)^{2(d-1)}}e^{i({\bf k}\cdot{\bf x}+{\bf k}'\cdot{\bf x}')}\\
\times\Big[\cos\omega_{\bf k}t\cos\omega_{\bf k'}t'\langle\varphi_L({\bf k},0)\varphi_L({\bf k}',0)\rangle_\chi
+\frac{\cos\omega_{\bf k}t\sin\omega_{\bf k'}t'}{\omega_{\bf k'}}\hspace{-0.5mm}\langle\varphi_L({\bf k},0)\pi_L({\bf k}',0)\rangle_\chi
\\
+\frac{\cos\omega_{\bf k'}t'\sin\omega_{\bf k}t}{\omega_{\bf k}}\hspace{-0.5mm}\langle\pi_L({\bf k},0)\varphi_L({\bf k}',0)\rangle_\chi
+\frac{\sin\omega_{\bf k}t\sin\omega_{\bf k'}t'}{\omega_{\bf k}\omega_{\bf k'}}\hspace{-0.5mm}\langle\pi_L({\bf k},0)\pi_L({\bf k}',0)\rangle_\chi\Big]\,,\end{multline}
where here and in the following $\langle \rangle_\chi$ denotes expectation values in the pre-quench vacuum $|0\rangle_\chi$.
Expanding the above correlators in terms of $\chi_\pm$, we obtain in terms of the Bogolyubov coefficients in \eqref{eq:bogcou} and \eqref{eq:bogcov},
\begin{align}
\label{eq:bilocal}
\langle T\{\varphi_L({\bf x},t)\varphi_L({\bf x}',t')\}\rangle_\chi 
=N\int\frac{\dd{\bf k}}{(2\pi)^{d-1}}&\frac{e^{i{\bf k}\cdot({\bf x}-{\bf x}')}}{2\omega_{\bf k}}\Big[\sum_\sigma u_{\bf k}^{L\sigma}v_{\bf k}^{L\sigma}  \cos\omega_{\bf k}(t+t')
\nonumber \\
&+\sum_\sigma (v_{\bf k}^{L\sigma})^2\cos\omega_{\bf k}(t-t')
+e^{-i\omega_{\bf k}|t-t'|} \Big]\,, 
\end{align}
At late times after the quench we can neglect the first term (the one with $(t+t')$ dependence) which oscillates with damped amplitude (cf. Appendix \ref{sect:stat_phase}). The remaining two terms correspond precisely to a thermal correlator with inverse temperature given by 
\begin{equation}
\beta_{\bf k}=\frac{1}{\omega_{\bf k}}\log{\left(\frac{\sum_\sigma (u_{\bf k}^{L\sigma})^2}{\sum_\sigma (v_{\bf k}^{L\sigma})^2}\right)}\,,
\end{equation}
in perfect agreement with \eqref{eq:temp_pred}.

\subsection{Smooth quenches} 
\label{subsect:smooth}
If one chooses to smoothen out $h(t)$ over a time scale $\Delta t$ then large momenta will be more suppressed, effectively cutting off the growth of the effective temperature at $|{\bf k}| \sim 1/\Delta t$. 
In fact, by representing our mixing quench as two mass quenches with subsequent field rotation, we can directly use classic results from quantum field theory on curved space time \cite{BirrellDavies1984}, as further developed in \cite{DasGalanteMyers2014,DasGalanteMyers2015}.
As mentioned in passing in Section \ref{sect:mixing} above, in the case of $m_L = m_R = m$, there is no a priori need to perform the $O(2)$ rotation in \eqref{eq:o2rotation} after the quench, since the action is already diagonal. Without it, the physical picture is that of two simultaneous mass quenches,
\begin{equation}
m_\pm^2(t) = m^2 \pm \theta(-t) h^2 \,.
\end{equation}
Now, instead of this instantaneous quench, we may consider a rapid, but smooth, quench of the form
\begin{equation}
m_\pm^2(t) = m^2 \pm \frac{h^2}{2} \left(1+\tanh\left(\frac{t}{\Delta t}\right)\right) \,,
\end{equation}
where $\Delta t$ is a predetermined (short) time scale. For this particular choice of time dependence, the analytical solutions to the mode functions that solve the corresponding Klein-Gordon equation, and thus the Bogolyubov coefficients relating incoming and outgoing modes are known explicitly \cite{BirrellDavies1984,DasGalanteMyers2014,DasGalanteMyers2015}.

Concretely, the field decomposition is similar to \eqref{eq:chi_decomp} except now the mode expansion also has a temporal profile,
\begin{equation}
\chi^\pm = \int\frac{\dd {\bf k}}{(2\pi)^{d-1}} e^{i{\bf k}\cdot {\bf x}}\left(f^\pm_{\bf k}(t)\, a_{{\bf k}}^{\pm} + {\left(f^\pm_{\bf k}(t)\right)^*}\, (a_{-{\bf k}}^{\pm})^{\dagger}\right)\,,
\end{equation}
with time dependent functions $f_k(t)$.
The incoming mode functions that correspond to plane waves at $t \to -\infty$ are expressed in terms of hypergeometric functions,
\begin{align}
f^\pm_{(in),{\bf k}} = \frac{1}{\sqrt{2\omega_{in}}}&\exp\left(-i \omega_p^\pm t - i \omega_m^\pm {\Delta t} \log\left(2\cosh(t/\Delta t)\right)\right)\nonumber \\
&\times {}_2F_1\left(1+i \omega_m^\pm {\Delta t}, i\omega_m^\pm {\Delta t}; 1- i\omega_{in}^\pm{\Delta t};\frac{1+\tanh(t/{\Delta t})}{2}\right)\,, \\
f^\pm_{(out),{\bf k}} = \frac{1}{\sqrt{2\omega_{out}}}&\exp\left(-i \omega_p^\pm t - i \omega_m^\pm {\Delta t} \log\left(2\cosh(t/\Delta t)\right)\right)\nonumber \\
&\times {}_2F_1\left(1+i \omega_m^\pm {\Delta t}, i\omega_m^\pm {\Delta t}; 1 + i\omega_{out}^\pm{\Delta t};\frac{1-\tanh(t/{\Delta t})}{2}\right)\,,\end{align}
where we have defined
\begin{align}
\omega^\pm_{in} &= \sqrt{|{\bf k}|^2 + m^2 \pm h^2}\,,\quad\, \omega_{out} = \sqrt{|{\bf k}|^2 + m^2}\,,\nonumber\\
\omega_p^\pm &= \frac{1}{2}(\omega_{out} + \omega_{in})\,,\qquad\quad \omega_m^\pm = \frac{1}{2}(\omega_{out} - \omega_{in})\,.
\end{align}
The two sets of mode functions are related by Bogolyubov transformations,
\begin{equation}
f^\pm_{(out),{\bf k}} = \mu_{\bf k}^\pm f^\pm_{(in),{\bf k}} + (\nu_{\bf k}^\pm)^* \left(f^\pm_{(in),{\bf k}}\right)^* \,,
\end{equation}
with coefficients
\begin{align}
\mu_{\bf k}^\pm &= \sqrt{\frac{\omega_{in}}{\omega_{out}^\pm}}\frac{\Gamma(1+i\omega_{out}^\pm{\Delta t})\Gamma(i \omega_{in}{\Delta t})}{\Gamma(1+i\omega_{p}^\pm{\Delta t})\Gamma(i \omega_{p}^\pm{\Delta t})} \,,\\
\nu_{\bf k}^\pm &= \sqrt{\frac{\omega_{in}}{\omega_{out}^\pm}}\frac{\Gamma(1-i\omega_{out}^\pm{\Delta t})\Gamma(i \omega_{in}{\Delta t})}{\Gamma(1-i\omega_{m}^\pm{\Delta t})\Gamma(-i \omega_{m}^\pm{\Delta t})} \,.
\end{align}
Combining this with the $O(2)$ rotation of the outgoing fields allows us to express the Bogolyubov transformation for the $L$ and $R$ fields, \eqref{eq:bogcou} and \eqref{eq:bogcov}, for a smooth quench,
\begin{equation}
\label{eq:bogcosmooth}
u_{\bf k} = 
\begin{pmatrix}
\mu_{\bf k}^+ &\quad -\mu_{\bf k}^- \\
\mu_{\bf k}^+ &\quad \mu_{\bf k}^-
\end{pmatrix}\,,\quad
{v}_{\bf k} = 
\begin{pmatrix}
\nu_{\bf k}^+ &\quad -\nu_{\bf k}^- \\
\nu_{\bf k}^+ &\quad \nu_{\bf k}^-
\end{pmatrix}\,.
\end{equation}
In the weak coupling limit, we obtain for the above coefficients
\begin{align}
\mu_{\bf k}^\pm &= 1 - \frac{h^4}{32 \omega_{out}^4} \Big(1+2\omega_{out}^2{\Delta t}^2 {\psi\left(i  \omega_{out}{\Delta t}\right)}\Big)\,,\\
\nu_{\bf k}^\pm &= 
\frac{\pi h^2  {\Delta t} \csch\left(\pi 
	\omega_{out}  {\Delta t}\right)}{8
\omega_{out}^{3}}\Big[\pm 2 \omega_{out}^2 +  i h^2 \omega_{out} {\Delta t} \Big( \psi\left(i \omega_{out}
{\Delta t}\right)+\gamma\Big) \Big]\,,
\end{align}
where $\psi(x)$ is the digamma function and $\gamma$ the Euler--Mascheroni constant.
We see that for $h \to 0$, all $v_{\bf k}$ vanish, as they should. Moreover, for $|{\bf k}| \gg 1/{\Delta t}$, the $v_{\bf k}$ are exponentially suppressed.

As an application, we calculate the effective temperature in the small $h$ limit, as per \eqref{eq:smallhtemp} and \eqref{eq:temperature}, where now $B_{\bf k}$ is complex but $|B_{\bf k}| \sim h^4$ as before.  We obtain
\begin{equation}
\beta_{\bf k} \approx \omega_{\bf k}^{-1}  \log \left(\frac{16 \omega_{\bf k}^2 \sinh ^2\left(\pi \omega_{\bf k} {\Delta t}\right)}{\pi ^2 {\Delta t}^2 h^4}\right)\,,
\end{equation}
which for large $k$ becomes
\begin{equation}
\beta_{\bf k} \approx 2\pi {\Delta t} + \frac{1}{|{\bf k}|}\log\left(\frac{4|{\bf k}|^2}{\pi^2 h^4 {\Delta t}^2}\right)\,.
\end{equation}
For fixed $h$, the temperature goes to a constant value at large $|{\bf k}|$ that is set purely by the time scale of the quench. This reflects the fact that modes with $|{\bf k}| \gg 1/{\Delta t}$ see an almost adiabatically changing mass and thus the probability of exciting them is exponentially small. Note also that for fixed $\Delta t$, the temperature goes to zero for $h \to 0$.

Let us end this section with a remark on the instantaneous limit of smooth quenches. It was noted in \cite{DasGalanteMyers2014} that in this limit, certain UV dominated quantities (e.g. $\langle \phi^2\rangle$) exhibit divergences at early times for $\delta t \to 0$, if the number of dimensions is large enough.
These are not vacuum divergences, but instead stem from the thermal parts of the propagators due to the $k$-dependent temperature. For any nonzero $\delta t$, $\beta_k$ goes to a constant for sufficiently large $k \gg \delta t^{-1}$. The Boltzmann factor thus provides a natural UV regulator. In stark contrast, in the case of an instantaneous quench, the Boltzmann factor only supplies a power law damping factor which fails to regulate UV divergences for a sufficiently high number of dimensions. We observe similar behavior for the spectral functions that we evaluate in the next section.

\section{Spacelike correlations in the post-quench background}\label{sect:dissipation}

In a non-vacuum background, such as a heat bath, correlation functions may exhibit behaviour that is forbidden in vacuum. 
Below, we illustrate how, even long after the quench, the non-trivial background is imprinted on the response of the system to small perturbations. 
In particular, modes which are exchanged with the background contribute to discontinuities of the retarded Green's function. We devote special attention to the regime $|\omega| \ll |{\bf k}|$, with $|{\bf k}|$ larger than any other scale. We can picture the damped spectral density we find in this parameter region intuitively in terms of an effective (approximate) thermalisation. Furthermore, there is an interesting parallel with strong coupling physics and dual gravity, discussed further in Section~\ref{sect:interpretation}. 

The spectral density function of the operator $\mathcal{O}({\bf x},t)$ in the state $\ket{\psi}$ is defined as
\begin{equation}\label{eq:spectral}
\mathcal{A}({\bf k},\omega) = -2 \Im[\mathcal{G}_{R}({\bf k},\omega)]\,,
\end{equation}
where $\mathcal{G}_{R}({\bf k},\omega)$ is the Fourier transform of the position-space retarded Green's function, which in turn is defined as
\begin{equation}\label{eq:retard}
G_{R}({\bf x},t) = -i\theta(t)\bra{\psi}[\mathcal{O}({\bf x},t),\mathcal{O}(0,0)]\ket{\psi}\,.
\end{equation}
In our case, $\ket{\psi}$ is the vacuum before the quench $\ket{0}_\chi$, {\it i.e.} the vacuum of the theory \eqref{eq:action} with constant $h$, and $\mathcal{O}({\bf x},t)=\normord{\varphi_L\varphi_L}\hspace{-1mm}({\bf x},t)$ is an $O(N)$ singlet operator, normal-ordered with respect to the physical vacuum after the quench, {\it i.e.} when the fields $\varphi_L$ and $\varphi_R$ diagonalize the action.

The definitions  \eqref{eq:spectral} and \eqref{eq:retard} rely on space- and time-translation invariance. More generally, $G_{R}$ is defined as
\begin{equation}\label{eq:retard1}
G_{R}({\bf x},{\bf x}',t,t') = -i\theta(t-t') \big\langle[\mathcal{O}({\bf x},t),\mathcal{O}({\bf x}',t')]\big\rangle_{\chi}\,,
\end{equation}
but in order to Fourier transform from the spatial coordinates to a single $(d-1)$-vector ${\bf k}$ and from the time coordinate to a single frequency $\omega$, we need $G_{R}$ to depend on only one $(d-1)$-vector ${\bf x}$ and one time-coordinate $t$. In the case at hand, time-translation invariance is explicitly broken by the quench but re-emerges at late times, as we will see below. Taking a Fourier transform of the resulting approximately time-translation invariant result leads to an effective $\mathcal{G}_{R}({\bf k},\omega)$ that makes physical sense for a late-time observer.

The computation of the position-space retarded Green's function in \eqref{eq:retard} is straightforward. The main steps are outlined in  Appendix \ref{app:retgf} and the end result is
\begin{align}
\label{GF-final}
G_R({\bf x},t) &= -i\theta(t)\big\langle[\normord{\varphi_L\varphi_L}\hspace{-1mm}({\bf x},t),\normord{\varphi_L\varphi_L}\hspace{-1mm}(0,0)]\big\rangle_{\chi}\notag\\
&= -i\theta(t) N\int\frac{\dd{\bf k}\dd{\bf k}'}{(2\pi)^{2(d-1)}}\frac{1}{4\omega_{\bf k}\omega_{\bf k'}}e^{-i({\bf k}-{\bf k}')\cdot {\bf x}}\notag\\
&\times\sum_\sigma\Big[\Big(1+(v^{L\sigma}_{\bf k})^2+(v^{L\sigma}_{\bf k'})^2\Big) \Big(e^{-i(\omega_{\bf k}+\omega_{\bf k'})t}-e^{i(\omega_{\bf k}+\omega_{\bf k'})t}\Big)
\notag\\
&+\Big((v^{L\sigma}_{\bf k'})^2-(v^{L\sigma}_{\bf k})^2\Big)\Big(e^{i(\omega_{\bf k}-\omega_{\bf k'})t}-e^{-i(\omega_{\bf k}-\omega_{\bf k'})t}\Big)
\Big]\,.
\end{align}
This can readily be Fourier transformed and extracting $\mathcal{A}({\bf k},\omega)$ according to \eqref{eq:spectral} yields
\begin{multline}\label{eq:symspectral}
\mathcal{A}({\bf k},\omega) = \pi N\int\frac{\dd{\bf k}'}{(2\pi)^{d-1}}\frac{1}{2\omega_{\bf k'}\omega_{{\bf k}-{\bf k}'}}\\
\times\sum_\sigma\Big[\Big(1+(v^{L\sigma}_{\bf k'})^2+(v^{L\sigma}_{{\bf k}-{\bf k}'})^2\Big)\Big(\delta(\omega_{\bf k'}+\omega_{{\bf k}-{\bf k}'}-\omega)-\delta(\omega_{\bf k'}+\omega_{{\bf k}-{\bf k}'}+\omega)\Big)
\\
+\Big((v^{L\sigma}_{\bf k'})^2-(v^{L\sigma}_{\bf k-k'})^2\Big)\Big(\delta(\omega_{\bf k'}-\omega_{{\bf k}-{\bf k}'}+\omega)-\delta(\omega_{\bf k'}-\omega_{{\bf k}-{\bf k}'}-\omega)\Big)
\Big]\,.
\end{multline}
As a consistency check, we note that ${\cal A}({\bf k},\omega)$ satisfies well-known positivity constraints that are a consequence of unitarity. In particular, $\omega\, {\cal A}({\bf k},\omega) > 0$, which is obvious for the terms on line two of \eqref{eq:symspectral} and for the terms on the bottom line it follows from the fact that $v_{\bf k}^{2}$ is a monotonically decreasing function of $|{\bf k}|$.

\begin{figure}[t!]
\centering
\setlength{\unitlength}{296.6220459bp}
	\begin{picture}(1,0.42015843)%
	\put(0,0){\includegraphics[width=\unitlength,page=1]{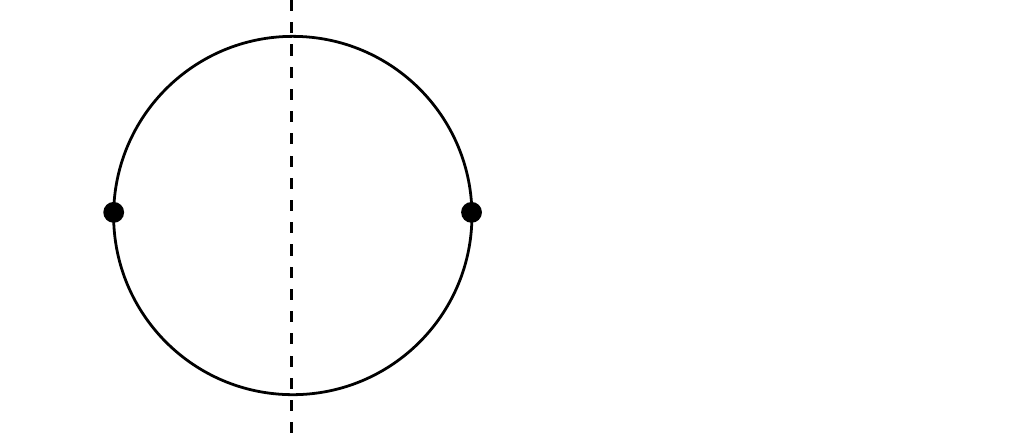}}%
	\put(0,0.20778947){\color[rgb]{0,0,0}\makebox(0,0)[lb]{\smash{\Large{Im}}}}%
	\put(0.55477887,0.20417731){\color[rgb]{0,0,0}\makebox(0,0)[lb]{\smash{\Large{$=$}}}}%
	\put(0,0){\includegraphics[width=\unitlength,page=2]{imdiagram.pdf}}%
	\put(0.97831047,0.37216368){\color[rgb]{0,0,0}\makebox(0,0)[lb]{\smash{2}}}%
	\put(0,0){\includegraphics[width=\unitlength,page=3]{imdiagram.pdf}}%
	\end{picture}
\caption{Extraction of imaginary part of Green's function using ordinary Cutkosky cutting rules. The on-shell propagators are thermal, with individual contributions to the right hand side illustrated in Fig.\ \ref{fig:diagrams}.}
\label{fig:imdiagram}
\end{figure}

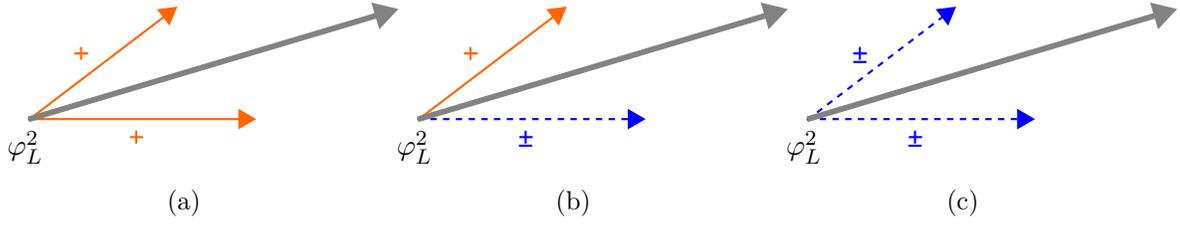
\begin{figure}[t]
\setlength{\unitlength}{146.89791539bp}
\centering
\vspace{5mm}
\hspace{-6.5mm}
\begin{subfigure}[b]{0.3\textwidth}
	\begin{picture}(1,0.39912949)%
	\put(0,0){\includegraphics[width=\unitlength,page=1]{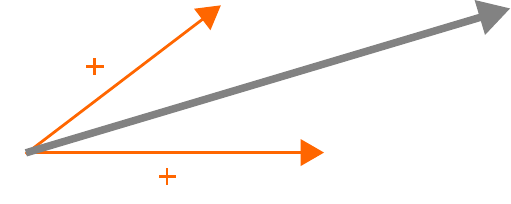}}%
	\put(-0.00284363,0.00788977){\color[rgb]{0,0,0}\makebox(0,0)[lb]{\smash{$\varphi_{L}^{2}$}}}%
	\end{picture}
	\vspace{-5mm}
	\caption{}
	\label{fig:diagrama}
\end{subfigure}
\quad 
\begin{subfigure}[b]{0.3\textwidth}
	\begin{picture}(1,0.39912949)%
	\put(0,0){\includegraphics[width=\unitlength,page=1]{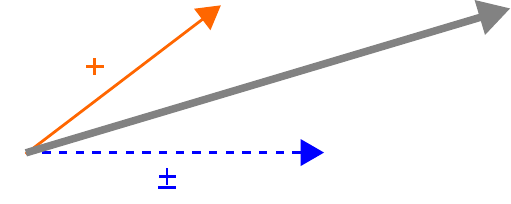}}%
	\put(-0.00284363,0.00788977){\color[rgb]{0,0,0}\makebox(0,0)[lb]{\smash{$\varphi_{L}^{2}$}}}%
	\end{picture}
	\vspace{-5mm}
	\caption{}
	\label{fig:diagramb}
\end{subfigure}
\quad
\begin{subfigure}[b]{0.3\textwidth}
	\begin{picture}(1,0.39912949)%
	\put(0,0){\includegraphics[width=\unitlength,page=1]{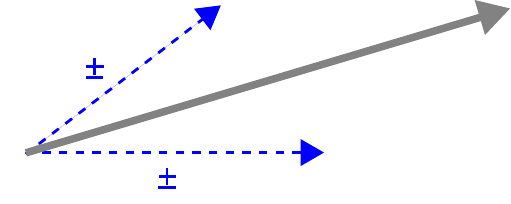}}%
	\put(-0.00284363,0.00788977){\color[rgb]{0,0,0}\makebox(0,0)[lb]{\smash{$\varphi_{L}^{2}$}}}%
	\end{picture}
	\vspace{-5mm}
	\caption{}
	\label{fig:diagramd}
\end{subfigure}
\caption{Different processes contributing to the spectral density.
	In all figures, the short dashed arrow is the momentum of a background excitation, the short solid arrow is the momentum of an external excitation, and the long arrow is 
	the sum of the two.
A plus sign represents a positive energy excitation whereas a minus sign represents energy absorption from the background. The process depicted in the final panel does not contribute in thermal equilibrium.
 }\label{fig:diagrams}
\end{figure}

A simple diagrammatic representation of the spectral density is given in Figs.\ \ref{fig:imdiagram} and \ref{fig:diagrams}.
${\cal A}$ corresponds to the imaginary part of the diagram depicted in Fig.\ \ref{fig:imdiagram}, obtained by cutting the loop and putting intermediate particles on shell. 
The terms on line two of \eqref{eq:symspectral} are due to on-shell creation of perturbations (Fig.\ \ref{fig:diagrama}, non-vanishing also in vacuum, i.e. for $h \to 0$), with an additional contribution from creating a perturbation while collectively exciting the background (Fig.\ \ref{fig:diagramb}). Both legs are either outgoing or incoming, whereby $\omega_{\bf k'}$ and $\omega_{{\bf k}-{\bf k}'}$ enter with the same sign.  The terms on the third line of \eqref{eq:symspectral}, on the other hand, correspond to processes where the momentum of the composite operator is absorbed by the background, while the total energy of the latter is lowered. Hence, one of the frequencies enters with a negative sign (Fig.\ \ref{fig:diagramb}). 

In the low energy regime $|\omega| \ll |{\bf k}|$, it is only the absorption terms on the bottom line of \eqref{eq:symspectral} that contribute, since the other delta functions cannot be satisfied. Taking $m^{2}/|{\bf k}|\ll \omega$ and inserting the Bogolyubov coefficients \eqref{eq:bog} we obtain, to leading order,
\begin{equation}
\label{eq:spect_evan}
\mathcal{A}({\bf k},\omega)\Big|_{|\omega| \ll |{\bf k}|} 
\simeq \frac{Nh^{4}}{2^{d+4}\pi^{d-2}}\int\dd|{\bf k}'|\frac{1}{|{\bf k}'|^{5}|{\bf k}-{\bf k}'|}(\delta(|{\bf k}'|-|{\bf k}-{\bf k}'|+\omega)-\delta(|{\bf k}'|-|{\bf k}-{\bf k}'|-\omega))\,,
\end{equation}
The integral can be evaluated in spherical coordinates.
Denoting the angle between ${\bf k}$ and ${\bf k'}$ by $\theta$, with $0 \leq \theta < \pi$ for $d > 3$, the arguments of the delta functions  are zero at
\begin{equation}
\theta_{\pm} = \arccos(\frac{|{\bf k}|^{2}\mp2|{\bf k}'|\omega -\omega^{2}}{2|{\bf k}||{\bf k}'|})\,.
\end{equation}
The solutions $\theta_{+}$ and $\theta_{-}$ only exist for
\begin{equation}
|{\bf k}'|\geq \frac{|{\bf k}|-\omega}{2}\,\qquad\textrm{and}\qquad |{\bf k}'|\geq \frac{|{\bf k}|+\omega}{2}\,,
\end{equation}
respectively,
which places lower limits on the range of integration over $|{\bf k}'|$. For $d = 3$, one has additional solutions at $-\theta_\pm$ because in this case $\theta \in [-\pi,\pi)$. Rewriting the delta functions using 
\begin{equation}
\delta\left(|{\bf k}'|-\sqrt{|{\bf k}|^{2}+|{\bf k}'|^{2}-2|{\bf k}||{\bf k}'|\cos\theta}\pm\omega\right) =
\frac{2(|{\bf k}'| \pm \omega)}{\sqrt{(|{\bf k}|^{2}-\omega^{2})((2|{\bf k}'| \pm \omega)^{2} - |{\bf k}|^{2})}}\delta(\theta - \theta_{\pm})\,,
\end{equation}
performing all angular integrals in \eqref{eq:spect_evan},
shifting $|{\bf k}'| \rightarrow |{\bf k}'| - \omega/2$ in the first term and $|{\bf k}'| \rightarrow |{\bf k}'| + \omega/2$ in the second, and neglecting higher order terms in $\omega$ we obtain
\begin{align}
\mathcal{A}({\bf k},\omega)\Big|_{|\omega| \ll |{\bf k}|} &= \frac{V_{d-2}'}{2^{2d-4}\pi^{d-2}} \frac{Nh^{4}\,\omega \,}{|{\bf k}|}\int_{\frac{|{\bf k}|}{2}}^{\infty}\dd |{\bf k}'|\,\frac{(4|{\bf k}'|^{2}-|{\bf k}|^{2})^{\frac{d-4}{2}}}{|{\bf k}'|^{5}}\notag\\
\label{eq:specfinal}
&= \frac{\Gamma\left(4-\frac{d}{2}\right)}{2^{2d-7}\pi^{\frac{d}{2}-1}}\frac{Nh^{4}\,\omega}{|{\bf k}|^{9-d}}\,,
\end{align}
where 
\begin{equation}\label{eq:norm}
V_{d-2}' = \frac{2\pi^{\frac{d}{2}-1}}{\Gamma\left(\frac{d}{2}-1\right)}\,,\quad\text{for}\quad d \geq 4\,,
\end{equation}
is the volume of the $(d-2)$-sphere divided by the contribution to this volume by the $\theta$-angle.

Equation \eqref{eq:specfinal} is our final expression for the spectral density in the low energy regime $|\omega| \ll |{\bf k}|$. It is valid for $d < 8$ since otherwise the dimensionless integral diverges
due to the growth of phase space compatible with
the on-shell condition in \eqref{eq:spect_evan} and the insufficient suppression of high momenta. 
As discussed in Section \ref{sect:density_operators}, however, this is caused by the instantaneous nature of the quench. We will return to this at the end of this subsection.
Furthermore,
one finds agreement between \eqref{eq:specfinal} evaluated at $d=2$ and an explicit computation in two dimensions. Hence, \eqref{eq:specfinal} is valid for $2 \leq d \leq 7$. 

We note that using the effective temperature \eqref{eq:temperature}, we can write \eqref{eq:specfinal} as
\begin{equation}\label{eq:specboltz}
\mathcal{A}({\bf k},\omega)\Big|_{|\omega| \ll |{\bf k}|} \sim \frac{N\omega}{|{\bf k}|^{5-d}}e^{-\beta_{{\bf k}/2}\,\frac{|{\bf k}|}{2}}\,,
\end{equation}
where the factor of $1/2$ in the exponent has been fixed by physical expectations, as the imaginary part of the retarded Green's function at low frequencies can be related to the absorption of an external perturbation with $\omega \approx 0$ and momentum ${\bf k}$ by a particle in the ``bath'' with frequency $\omega = \frac{|{\bf k}|}{2}$ and momentum $-\frac{{\bf k}}{2}$ while the latter remains on-shell \cite{SonStarinets2002}. As expected, we find agreement with free thermal field theory in the high temperature phase \cite{AmadoSundborgThorlaciusWintergerst2018}.

Finally, if we consider a smooth but rapid quench rather than an instantaneous one, then according to subsection \ref{subsect:smooth}, the temperature is bounded by the time scale of the quench, $\beta_{\bf k} \to \beta_\infty \equiv 2\pi\Delta t$ for $|{\bf k}| \to \infty$. This has two important consequences. First, the integral in \eqref{eq:specfinal} is rendered finite due to the exponential suppression of contributions from large $|{\bf k}'|$, and second, the lower integration limit directly provides a Boltzmann factor $e^{-\beta_\infty \frac{|{\bf k}|}{2}}$, leading to perfect agreement with \cite{AmadoSundborgThorlaciusWintergerst2018}.

The support for momenta outside the light-cone in the absorptive part of the response function means that the background acts as source and sink of composite operators with such momenta. Although these operators are formally local, their constituents have non-trivial momentum distributions due to the background. In effect, the composite operators inherit a relative position distribution, with a scale  determined by a characteristic wavelength. In the background they function as extended objects whose size enters their correlation functions. This argument does not really depend on interaction strength and we are led to expect qualitatively similar `non-localities' at weak and strong coupling. Another perspective on the importance of compositeness is interference: the spectral density \eqref{eq:symspectral} includes `beats', {\it i.e.} terms with frequency equal to the difference of frequencies of the constituent fields. Interference with modes in the background is directly detectable in composite operators. In the particle picture, constituents are exchanged with the background and it is crucial that energy quanta can be extracted from the background.

\section{The holographic perspective} \label{sect:interpretation}

All quantum field theory results of the previous sections can be given a holographic interpretation. It works for large $N$ quantum field theories with scale symmetric high energy limits, in particular for free massive $O(N)$ models whose high energy limits are the corresponding massless models. We now explain this interpretation and argue that the quench setup permits a realisation of the proposed ER = EPR relation \cite{MaldacenaSusskind2013}. It states that Einstein-Rosen bridges, which are inside black hole horizons and connect two different regions of asymptotic spacetime, are dual to Einstein-Podolsky-Rosen entanglement of two corresponding quantum subsystems. We have constructed entangled states of two quantum field theories from a quench \`a la Einstein-Podolsky-Rosen. The present section first explains our perspective on holography, then how black holes can be detected in general, how to probe the geometries from the boundary, and finally how it can be seen that there is a bridge between two asymptotic regions of spacetime.

Assuming that quantum field theories can be decoded as bulk gravitational physics \cite{Witten1998} has led to many insights and new ideas. It can even be taken to define bulk physics. 
In the following we will probe bulk physics operationally from its detection at the boundary. For quantum field theories with an ultraviolet fixed point, like the model we consider, the holographic dictionary is unambiguous. The fixed point corresponds to an asymptotically AdS region in which spins and masses of bulk fields are given by the spins and scaling dimensions of boundary operators. What happens deep inside the bulk depends on a background configuration of bulk fields and on interactions between the bulk modes corresponding to boundary operators. The interpretation is then organised by the expansion parameter $1/N$ in large $N$ gauge theories or $O(N)$ models.

Although we wish to emphasise general holography rather than a particular model, there is one proposed and partially tested holographic duality, which is closely related to our $O(N)$ model setup. After the initial connection between AdS higher spin symmetry and holography \cite{Sundborg2001,Witten2001,SezginSundell2002} the massless free and critical $O(N)$ models were proposed \cite{KlebanovPolyakov2002} as boundary duals of pure Fradkin-Vasiliev higher spin theories \cite{FradkinVasiliev1987,FradkinVasiliev1987a,FradkinVasiliev1987b}. For a review of these theories, their holography and tests of it, see \cite{Giombi2016}. For our purposes it is enough to know that we are working with a massive (infrared deformed) version of a model dual to a (higher spin) gravitational theory, thus a geometry that deviates from AdS in the interior, but not asymptotically. It is thus an example of the framework sketched above. Spins and masses of bulk fields are given by the spins and scaling dimensions of boundary operators and the interpretation is organised by the expansion parameter $1/N$.

\subsection{Horizon detection}

In order to establish a holographic interpretation of our results, let us first review some basic properties of horizons as seen from the boundary in semiclassical gravity on AdS. Observing analogous signatures in our weakly coupled boundary theories will then allow us to draw an approximate holographic picture.

Black hole-like objects can be detected by scattering waves against the object and looking at the size of its shadow. In asymptotically AdS geometries scattering and wave detection is represented by boundary correlation functions, which can be obtained in quantum field theory according to AdS/CFT. We have chosen 2-point correlation functions of scalar operators corresponding to bulk scalar fields. Now, for the interpretation we need to know the signal from scattering on an AdS black hole. 

\begin{figure}[t!]
	\setlength{\unitlength}{176.85823623bp}
	\centering
	\vspace{5mm}
	\begin{subfigure}[b]{0.4\textwidth}
		\begin{picture}(1,0.60301526)%
		\put(0,0){\includegraphics[width=\unitlength,page=1]{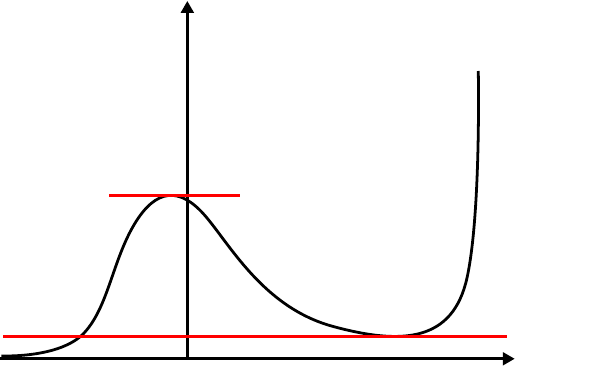}}%
		\put(0.22305808,0.57902979){\color[rgb]{0,0,0}\makebox(0,0)[lb]{\smash{$V$}}}%
		\put(0.86704711,0.00514855){\color[rgb]{0,0,0}\makebox(0,0)[lb]{\smash{$r_{*}$}}}%
		\end{picture}
		\vspace{-5mm}
		\caption{}
		\label{fig:potentialSch}
	\end{subfigure}
	\qquad 
	\begin{subfigure}[b]{0.4\textwidth}
		\begin{picture}(1,0.60301526)%
		\put(0,0){\includegraphics[width=\unitlength,page=1]{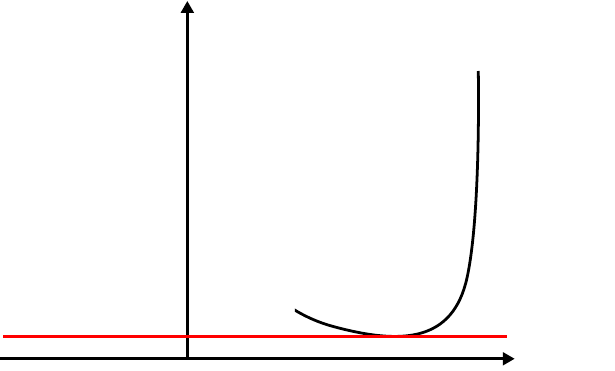}}%
		\put(0.22305808,0.57902979){\color[rgb]{0,0,0}\makebox(0,0)[lb]{\smash{$V$}}}%
		\put(0.86704711,0.00514855){\color[rgb]{0,0,0}\makebox(0,0)[lb]{\smash{$r_{*}$}}}%
		\put(0,0){\includegraphics[width=\unitlength,page=2]{potentialAdS.pdf}}%
		\end{picture}
		\vspace{-5mm}
		\caption{}
		\label{fig:potentialAdS}
	\end{subfigure}
	\caption{Depiction of the effective Schrödinger potential in (a) Schwarzschild--AdS, and (b) Global AdS. The lower red line corresponds to the global minimum in pure AdS whereas the upper red line in (a) corresponds to the local maximum of the angular momentum barrier.}\label{fig:barrier}
\end{figure}

The propagation of scalars in an AdS black hole background is encoded in its mode spectrum, which depends on the radial wave equation in the geometry. The wave equation, in its turn, is affected by the geometry through an effective potential $V_{\textrm{eff}}$ including an important term from angular motion, which produces an angular momentum barrier (Figure \ref{fig:barrier}). The results described in Section~\ref{sect:dissipation} have an interesting counterpart in so-called evanescent modes in gravitational theories dual to strongly coupled gauge theories \cite{ReyRosenhaus2014}. We review in Appendix \ref{sect:evanescence} how the effective potential is obtained and the definition of the convenient radial ``tortoise coordinate''. Here we focus on the evanescent modes.

Two properties are now important for our discussion: 
\begin{itemize}
	\item If there are modes with \emph{frequency lower than the minimum allowed by the AdS background} matching the geometry asymptotically, they are asymptotically suppressed relative to the AdS modes.
	\item When the angular momentum barrier is responsible  for the asymptotic suppression, the damping is stronger with increasing angular momentum. The damping is exponential in momentum when, as is typical, the WKB approximation is applicable.
\end{itemize}
We refer to modes satisfying the above criteria as evanescent, because of their falloff in the radial direction, perpendicular to their propagation in angular directions. As a rule of thumb, they appear if the effective potential has a sufficiently deep and wide well inside the angular momentum barrier. In particular, this happens for AdS black holes, where the well extends to the horizon at $r =-\infty$ in tortoise coordinates. It has also been argued that stars of ordinary matter are unlikely to be small enough to have the requisite well in their effective potential \cite{ReyRosenhaus2014}. 

To connect to the flat space quantum field theory calculation we now take a large radius limit. The curvature of the fixed radius surfaces vanishes in this limit, angles go to flat coordinates and angular momentum goes to momentum after rescaling. Evanescent modes in the resulting planar geometry are then characterised by spacelike wave-vectors in the transverse space, due to the anomalously low frequency mentioned above. The low frequency is due to the strongly attractive nature of the effective potential.

\subsection{Horizon detection from quantum field theory}

We are now ready to interpret our quench results holographically. Since the quench injects energy into the system, the post-quench state is characterised by a finite expectation value of the stress energy tensor. According to the standard holographic dictionary, this implies deviations of the bulk metric from empty AdS to some asymptotically AdS spacetime. 

To obtain basic information on the deviation of the geometry from AdS, we can probe it with bulk fields. The simplest bulk probe is dual to the boundary operator $\normord{\varphi^2}$. We calculated its response function in the field theory in section \ref{sect:dissipation} and found support for its absorptive part at spacelike momenta. This detects absorption. Furthermore, the damping  of large momenta in the absorption becomes exponential if the instantaneous quench is smoothed out, Such behaviour is in fact familiar from the evanescent modes discussed in the previous horizon detection section. There, the absorptive part means absorption at the horizon. It comes about because gravitational attraction wins against an effective (angular) momentum repulsion in the deep interior horizon region. The momentum suppression is crucial to this identification, since it indicates that bulk modes have to tunnel through the momentum barrier in order to get absorbed. To summarize, a standard probe interpretation of the boundary correlation function conforms qualitatively with boundary correlator probes in a classical geometry with a black hole horizon.

Wave equations require specification of boundary conditions at horizons and waves which are absorbed or emitted from a horizon correspond to dissipation with a continuous spectrum. If, on the other hand, the bulk spacetime has no horizon, 
then the spectrum will in general be discrete. Thus, in so far as the evanescent mode spectrum of our field theoretic system is continuous, as demonstrated by the spectral density (\ref{eq:specfinal}), a holographic interpretation includes a horizon. 

\subsection{Bridge over censored regions}

The after-quench system can be viewed entirely in terms of the $L$ degrees of freedom to 
describe a single sided bulk space time with an effective horizon.
The $R$ degrees of freedom yields a mirror image of the black hole. When $L$ operators exclusively, or $R$ operators exclusively, are used to probe the system, the effect is identical to tracing over the other Hilbert space, effectively working in mixed ensembles resembling thermal ensembles, as demonstrated in previous sections. The $L, R$ system of independent quantum fields represents the two independent boundaries. Just as in the eternal black hole \cite{Maldacena2003}, $L$ and $R$ fields are correlated only by an entangled state, here resulting from the quench. The non-zero correlation, 
is crucial: it guarantees that bulk geometry is connected. We conclude that we have a connected $L$-$R$-symmetric geometry with two asymptotic regions and two horizons, much like a standard eternal black hole.

In the quench, the microscopic $O(N)_L$ and $O(N)_R$ degrees of freedom are correlated due to the mixing in the pre-quench Hamiltonian, which prepares the state. The free $O(N)_L\times O(N)_R$ symmetry is then broken to $O(N)_\text{diag}$. This symmetry breaking is in common with the high temperature phase of $O(N)$ singlet models \cite{JevickiYoon2016,AmadoSundborgThorlaciusWintergerst2018}. 

\section{Conclusions}\label{sect:conclusions}

In this paper we have considered a pair of free field theories in a highly entangled state, generated by the instantaneous quench of a mixing interaction between the fields. Probing the entangled system long after the quench, using operators restricted to a single field theory, reveals an approximately thermal background at long wavelengths. The similarity to a thermal system becomes more pronounced when the quench is described by a smooth but rapidly varying coupling rather than an instantaneous jump.  

Interesting parallels with strongly coupled theories and holographic duality emerge when the post-quench state is probed by composite operators, rather than just the elementary fields. Correlations between quadratic composite operators remain long after oscillations induced by the quench have subsided. This includes, in particular, correlation functions that are damped but non-vanishing at spacelike momenta, in contrast to correlation functions of elementary excitations in the post-quench background or indeed correlation functions of the same composite operators in the vacuum. 

Analogous behaviour is seen in well established gravitational duals of strongly coupled gauge theories \cite{ReyRosenhaus2014}. There one considers boundary-to-boundary correlation functions in asymptotically AdS geometries with localised bulk gravitational sources (typically black holes) and finds non-vanishing correlations at spacelike momenta. On the gravity side, these correlations are associated with so-called evanescent modes, modes that are radially localized near the bulk source and exponentially suppressed towards the boundary.  

Our results suggest that this a general phenomenon associated with composite operators in non-trivial backgrounds, independent of coupling strength. Indeed, similar behaviour has been observed in thermal free field theory \cite{JevickiYoon2016,JevickiSuzuki2016,AmadoSundborgThorlaciusWintergerst2018} and there are general arguments for a thermal field theory bound on the size of spacelike correlators at large momenta \cite{BanerjeePapadodimasRajuSamantrayShrivastava2019}, for any coupling. More generally, composite operators are sensitive to non-trivial backgrounds, both thermal and non-thermal, at both weak and strong coupling.

\begin{acknowledgments}
The work of BS was supported in part by the Swedish Research Council under 
contract DNR-2018-03803. The work of LT was supported in part by the Icelandic Research Fund grant 195970-051 and the University of Iceland Research Fund. NW acknowledges support by FNU grant
number DFF-6108-00340. The work of SB is supported by the Knut and Alice Wallenberg Foundation under grant 113410212. The work was finished while BS was participating in a programme at ESI, the Erwin Schr\"odinger International Institute in Vienna, whose stimulating environment is also acknowledged.
\end{acknowledgments}

\begin{appendix}

\section{Effective potential for scalar waves around black holes}
\label{sect:evanescence}
Following \cite{ReyRosenhaus2014} we
consider a minimally coupled scalar field $\Phi$ on a static spherically symmetric geometry described by the line element
\begin{equation}
\dd s^{2}= - f(r)\dd t^{2} +\frac{\dd r^{2}}{f(r)}+r^{2}\dd \Omega^{2}_{d-1}\ .
\end{equation}
The Klein-Gordon equation in this background can be solved by separation of variables, $\Phi(t,\Omega,r)=e^{-i\omega t}Y(\Omega)\phi(r)$, and with $d$ denoting the spatial dimension, the radial equation takes the form
\begin{equation}
0 = \frac{1}{r^{d-1}}\partial_{r}\left(f\ r^{d-1}\partial_{r}\phi\right) +\left( \frac{\omega^{2}}{f} - \frac{l(l+d-2)}{r^{2}}-m^{2}\right)\phi\ ,
\end{equation}
which is conveniently rewritten by replacing $\phi(r)=u(r)/r^{\frac{d-1}{2}}$ and using ``tortoise'' coordinates, $\dd r_{*}=\dd r/f$:
\begin{equation}
0= \frac{\dd ^{2}u}{\dd r_{*}^{2}} + \left\{\omega^{2}-V(r(r_{*}))\right\}\,u\ ,
\end{equation}
where the effective potential $V$ is given by
\begin{equation}
V(r)= \left[ \frac{l(l+d-2)}{r^{2}} + \frac{(d-1)(d-3)f(r)}{4r^{2}} + \frac{(d-1)f'(r)}{2r} + m^{2}\right] f(r)\ .
\end{equation}
The advantage of these coordinates is that we recognise the form of the radial time-independent Schr\"odinger equation, and can invoke our intuition about its solutions. 

\section{Stationary phase approximation}
\label{sect:stat_phase}
Since the state after the quench is not time independent, correlation functions are not bound by time translational invariance. In particular, two-point functions of the form $\langle{\cal O}(t) {\cal O}(t')\rangle$ will acquire an explicit dependence on $t+t'$. However, as we demonstrate here, for a large class of operators these contributions decay sufficiently fast at late times, thus implying an emergent time-translational invariance.

The most general form of a factor depending on $(t+t')$ that appears in relevant expressions is 
\begin{align}
\label{eq:statph}
\int\dd{\bf k}f({\bf k})\,e^{\pm i (\omega_{\bf k}(t+t')-{\bf k}\cdot{\bf x})} \,.
\end{align}
Now, $\omega_{\bf k}$ has a global minimum at ${\bf k} =0$ which implies that the main contribution to the integral for large $t+t'$ and fixed ${\bf x}$ will be around this point. Thus the integral can be approximated by that over an $\epsilon$-ball around ${\bf k} = 0$,
\begin{align}
\int_{\epsilon} \dd{\bf k}f({\bf k})\,e^{\pm i (\omega_{\bf k}(t+t')-{\bf k}\cdot{\bf x})} &\simeq \int_{\epsilon} \dd{\bf k}f({\bf k})\,e^{\pm i \left(\left(m + \frac{|{\bf k}|^2}{2m}\right)(t+t')-{\bf k}\cdot{\bf x}\right)} \notag\\
&\simeq f(0)\,e^{\pm i m (t+t')}\int_{\epsilon} \dd{\bf k}\,e^{\pm i \frac{|{\bf k}|^2}{2m}(t+t')} \notag\\
& \simeq f(0)\left(\frac{2m}{t+t'}\right)^{\frac{d-1}{2}}\,e^{\pm i m (t+t')}\int \dd{\bf y}\,e^{\pm i y^2}\,, 
\end{align}
where in the last step we have formally extended the region of integration back to $\mathbb{R}^{d-1}$, since for large $(t+t')$ the two integrals agree up to exponentially small corrections. The final integral can be easily verified to be finite, whereby we see that for $d>1$ the integral \eqref{eq:statph} will go to zero at late times. 

\section{The retarded Green's function} \label{app:retgf}
In order to compute the retarded Green's function we need the field operator, $\varphi_L({\bf x},t)$ which assumes a mode expansion of the form 
\begin{equation}
\varphi_L({\bf x},t) = e^{-i H t} \varphi_L e^{i H t}  = \int\frac{\dd{\bf k}}{(2\pi)^{d-1}}\frac{1}{\sqrt{2\omega_{\bf k}}}\left(L_{{\bf k}}e^{-ik\cdot x} + (L_{{\bf k}})^{\dagger}e^{ik\cdot x}\right)\,,
\end{equation}
with $k\cdot x\equiv \omega_{\bf k} t - {\bf k\cdot x}$ and $\varphi_L$ given in \eqref{eq:phi_decomp}.

Our aim is to compute the retarded Green's function, \eqref{eq:retard1} at a late time, $t$ for the $O(N)$ singlet operators in the pre-quench vacuum characterized by non-vanishing coupling, $h$ : 
\begin{equation}
\label{GF1}
G_{R}({\bf x},{\bf x}',t,t') = -i\theta(t-t') \, \big\langle\left[\normord{\varphi_L\varphi_L}\hspace{-1mm}({\bf x},t),\normord{\varphi_L\varphi_L}\hspace{-1mm}({\bf x}',t')\right]\big\rangle_\chi.
\end{equation}

Using the Bogolyubov transformation defined in \eqref{eq:bog} the normal ordered singlet operator can be expanded as
\begin{multline}
\normord{\varphi_L({\bf x},t)\varphi_L({\bf x},t)} = \\
\int\frac{\dd{\bf k}\,\dd{\bf k}'}{(2\pi)^{2(d-1)}}\frac{1}{\sqrt{4\omega_{\bf k}\omega_{\bf k'}}}
\Big( L_{\bf k} L_{\bf k'} e^{-i \left(k + k'\right)\cdot x} +  L_{\bf k}^\dagger L_{\bf k'}^\dagger e^{i \left(k + k'\right)\cdot x} \\
+ L_{\bf k}^\dagger L_{\bf k'} e^{i \left(k - k'\right)\cdot x} + L_{\bf k'}^\dagger L_{\bf k} e^{i \left(k' - k\right)\cdot x}\Big),
\end{multline}
with 
\begin{equation}
L_{\bf k} = \frac{1}{\sqrt{2}}\sum_{\sigma} u^{L\sigma}_{\bf k} a^{\sigma}_{\bf k} + v^{L\sigma}_{\bf k}(a^{\sigma}_{-{\bf k}})^{\dagger}\,.
\end{equation}

As an intermediate step, let us investigate the individual terms in \eqref{GF1}, namely the two point correlator of the singlet field operator,
\begin{equation}
\label{corr1}
{\cal C}({\bf x},{\bf x}',t,t') = \big\langle\normord{\varphi_L\varphi_L}\hspace{-1mm}({\bf x},t)\normord{\varphi_L\varphi_L}\hspace{-1mm}({\bf x}',t')\big\rangle_\chi\,.
\end{equation}
Quite a lot of simplifications are already in order at this level:
\begin{itemize}
\item Each term will contain a product of four creation and annihilation operators sandwiched between the pre-quench vacuum states. Only number-preserving permutations among these would give non-vanishing contributions to \eqref{corr1}.
\item The summation over $O(N)$ indices will give rise to different scaling behaviours with powers of $N$ for different terms in the expansion of \eqref{corr1}. It can be readily verified using the canonical commutation relations of the creation and annihilation operators that the terms either scale as $N$ or as $N^2$. For instance, while the terms containing the permutations, $\langle a_{{\bf k}}^{-}a^{-}_{{\bf k}'}(a_{{\bf k}''}^{-})^{\dagger} (a_{{\bf k}'''}^{-})^{\dagger}\rangle_\chi$ and $\langle a_{{\bf k}}^{-}a^{+}_{{\bf k}'}(a_{{\bf k}''}^{-})^{\dagger}(a_{{\bf k}'''}^{+})^{\dagger}\rangle_\chi$ scale as $N$, the terms consisting of  $\langle a_{{\bf k}}^{-}(a^{-}_{{\bf k}'})^{\dagger}a_{{\bf k}''}^{-} (a_{{\bf k}'''}^{-})^{\dagger}\rangle_\chi$ and $\langle a_{{\bf k}}^{+}(a^{+}_{{\bf k}'})^{\dagger}a_{{\bf k}''}^{-} (a_{{\bf k}'''}^{-})^{\dagger}\rangle_\chi$ scale as $N^2$.
\item However, since the total ${\cal O}(N^2)$ contribution to \eqref{corr1} just corresponds to the disconnected piece of the Green's function, it is necessarily symmetric under $x \leftrightarrow x'$. Such terms are therefore guaranteed to give vanishing contributions to the retarded Green's function \eqref{GF1} by definition. Therefore for our purpose, it is sufficient to consider terms which give ${\cal O}(N)$ contributions to \eqref{corr1}.
\item For the same reason, terms which are only functions of $(x+x')$ can be ignored.
\item Last but not least, invoking the results from appendix \ref{sect:stat_phase}, all terms containing $e^{\pm i \omega_{\bf k} (t+t')}$ can be rendered negligible as far as their late time behaviours are concerned. This yields an emergent time-translation invariance at late time which allows us to set $t' = 0$ for all practical purposes in view of the fact that ultimately we are interested in evaluating the Green's function at late times.
\end{itemize}

These steps lead to a simple expression for the Green's function, displayed also in the main text as equation \eqref{GF-final}.
\begin{align}
G_R({\bf x},t) &= -i\theta(t)\big\langle[\normord{\varphi_L\varphi_L}\hspace{-1mm}({\bf x},t),\normord{\varphi_L\varphi_L}\hspace{-1mm}(0,0)]\big\rangle_\chi\notag\\
&= -i\theta(t) N\int\frac{\dd{\bf k}\dd{\bf k}'}{(2\pi)^{2(d-1)}}\frac{1}{4\omega_{\bf k}\omega_{\bf k'}}e^{-i({\bf k}-{\bf k}')\cdot {\bf x}}\notag\\
&\times\sum_\sigma\Big[\Big(1+(v^{L\sigma}_{\bf k})^2+(v^{L\sigma}_{\bf k'})^2\Big) \Big(e^{-i(\omega_{\bf k}+\omega_{\bf k'})t}-e^{i(\omega_{\bf k}+\omega_{\bf k'})t}\Big)
\notag\\
&+\Big((v^{L\sigma}_{\bf k'})^2-(v^{L\sigma}_{\bf k})^2\Big)\Big(e^{i(\omega_{\bf k}-\omega_{\bf k'})t}-e^{-i(\omega_{\bf k}-\omega_{\bf k'})t}\Big)
\Big]\,.
\end{align}

\end{appendix}

\bibliographystyle{utphys}
\bibliography{Bibliography}

\end{document}